\documentclass[english]{iopart}
\usepackage{mathptmx}
\usepackage{helvet}
\usepackage{courier}
\usepackage[T1]{fontenc}
\usepackage[latin9]{inputenc}
\usepackage[a4paper]{geometry}
\usepackage{units}
\usepackage{iopams}
\usepackage{amssymb}
\usepackage{graphicx}
\makeatletter

\@ifundefined{textcolor}{}
{%
 \definecolor{BLACK}{gray}{0}
 \definecolor{WHITE}{gray}{1}
 \definecolor{RED}{rgb}{1,0,0}
 \definecolor{GREEN}{rgb}{0,1,0}
 \definecolor{BLUE}{rgb}{0,0,1}
 \definecolor{CYAN}{cmyk}{1,0,0,0}
 \definecolor{MAGENTA}{cmyk}{0,1,0,0}
 \definecolor{YELLOW}{cmyk}{0,0,1,0}
 }

\usepackage{graphicx}

\makeatother

\usepackage{babel}
\begin{document}

\title{Symmetries of cyclic work distributions for an isolated harmonic
oscillator}

\author{Ian J. Ford, David S. Minor and Simon J. Binnie }

\address{Department of Physics and Astronomy and London Centre for Nanotechnology,
University College London, Gower Street, London WC1E 6BT, United Kingdom}
\ead{i.ford@ucl.ac.uk}
\begin{abstract}
We have calculated the distribution of work $W$ done on a 1-d harmonic
oscillator that is initially in canonical equilibrium at temperature
$T$, then thermally isolated and driven by an arbitrary time-dependent
cyclic spring constant $\kappa(t)$, and demonstrated that it satisfies
$P(W)=\exp(\beta W)P(-W)$, where $\beta=1/k_{B}T$, in both classical
and quantum dynamics. This differs from the celebrated Crooks relation
of nonequilibrium thermodynamics, since the latter relates distributions
for forward and backward protocols of driving. We show that it is
a special case of a symmetry that holds for non-cyclic work processes
on the isolated oscillator, and that consideration of time reversal
invariance shows it to be consistent with the Crooks relation. We
have verified that the symmetry holds in both classical and quantum
treatments of the dynamics, but that inherent uncertainty in the latter
case leads to greater fluctuations in work performed for a given process.
\end{abstract}
\maketitle

\section{Introduction}

The Crooks relation \cite{Crooks99} states that the outcome of the
mechanical processing of a system according to a prescribed sequence
of actions is related to the outcome of a process consisting of the
reversed sequence. It is a connection between the probability distributions
of the amount of work $W$ performed on the system in the course of
such forward and backward processes, $P_{\rm F}(W)$ and $P_{\rm B}(W)$,
respectively. A forward process might consist of the movement of a
piston to compress a gas in a cylinder, while the backward process
would be the opposite movement to expand the gas \cite{Crooks07}.
The validity of the Crooks relation requires that the system should
start out in canonical equilibrium at the same temperature $T$ for
both processes. The system might maintain contact with a heat bath
at that temperature during the processing or it could be isolated.
The relation reads
\begin{equation}
P_{\rm F}(W)=\exp\left(\beta\left(W-\Delta F\right)\right)P_{\rm B}(-W),\label{1}
\end{equation}
where $\beta=1/k_BT$ and $\Delta F$ is the change in free energy of
the system associated with the forward process, evaluated for example
on the basis of the isothermal change in volume of the expanded gas.
Equation (\ref{1}) states that the probability that the forward process
should require an input of work $W$, and the probability of requiring
work $-W$ (in other words receiving work from the system) during
the backward process, are related to each other, but are not in general
equal. The relation has been shown to hold for a variety of choices
of dynamics, though studies reveal that it is important to define
carefully what is meant by work, particularly for strong coupling
between a system and its environment \cite{Campisi11,Jarstrong}.
It implies \cite{Jarzynski97} the Jarzynski equality $\langle\exp\left(-\beta\left(W-\Delta F\right)\right)\rangle=1$,
which in turn leads to $\langle W\rangle\ge\Delta F$ for a forward
process starting in equilibrium, where the brackets indicate an average
taken over the probability distribution of work done. This, of course,
is a statement of the second law of thermodynamics, and the Crooks
relation, Jarzynski equality and the associated fluctuation relations
have received a great deal of attention as a result (see, for example
\cite{Evans93,Evans94,Bochkov81,Gallavotti95,jaroriginal1,kurchan,Lebowitz99,Evans02,Jarpathintegral,Harris07,Crooks08,Esposito09,Campisi11,SpinneyFordChapter12}).

It has proved valuable to study the Crooks relation in the context
of simple examples \cite{Imparato05,Crooks07,Talkner08,Deffner08,Saha09,vanZon08a,Huber08}
in order to gain insight into its operation in more complex cases,
and the 1-d harmonic oscillator has proved to be a popular system.
The main purpose of this paper is to calculate work distributions
for an isolated oscillator using a geometric and pictorial approach
that has, we believe, some intuitive pedagogical value. In doing so,
we expose a broader symmetry of the distribution of work for such
a forward process, which has its origin in the simplicity of the dynamics
of the harmonic oscillator, and which demonstrates the rather special
character of this system with regard to its fluctuation behaviour.

Work is performed by prescribing a time-dependent spring constant
$\kappa(t)$ during the process. If the spring constant varies cyclically
in an interval $0\le t\le\tau$ such that $\kappa(\tau)=\kappa(0)$,
then $\Delta F=0$, and moreover if the process takes place under
conditions of thermal isolation, then no heat is exchanged during
the cycle and $W=\Delta E$, the change in system energy. The Crooks
relation reduces to $P_{\rm F}(\Delta E)=\exp(\beta\Delta E)P_{\rm B}(-\Delta E)$:
a result verified, for example, by Deffner and Lutz \cite{Deffner08,Deffner10}.
But our approach to solving the classical evolution demonstrates that
a relation $P_{\rm F}(\Delta E)=\exp(\beta L\Delta E)P_{\rm F}(-\Delta E)$
also holds for a class of forward processes, with the parameter $L$
depending on the nature of the process, and with $P_{\rm F}(\Delta E)$
taking a specified analytic form. It is quite compatible with the
Crooks relation for a cyclic forward process and its backward counterpart,
for which $L=1$ and $P_{\rm F}(\Delta E)=P_{\rm B}(\Delta E)$, as we shall
show. We also compare the continuous distribution of work arising
from a classical treatment with the discrete distribution of work
that emerges from the quantum treatment of a cyclic process to show
that these properties are preserved. Fluctuations in the quantum case
are broader, as a result of the wider range of possible outcomes made
possible by the dynamics.

In the next section we analyse a general work process performed on
a classical harmonic oscillator, represented in terms of a matrix
operation on a system phasor. We evaluate the probability distribution
function (pdf) of system energy change for $M\ge 1$ independent oscillators.
We go on to treat the system quantum mechanically in section \ref{sec:Quantum-treatment},
particularly to contrast the widths of the classical and quantum pdfs
for cyclic processes. Our conclusions are given in section \ref{sec:Conclusions}.

\section{Classical treatment}

\subsection{System phasor and process matrix}

\begin{figure}
\begin{center}\includegraphics[width=0.7\columnwidth]{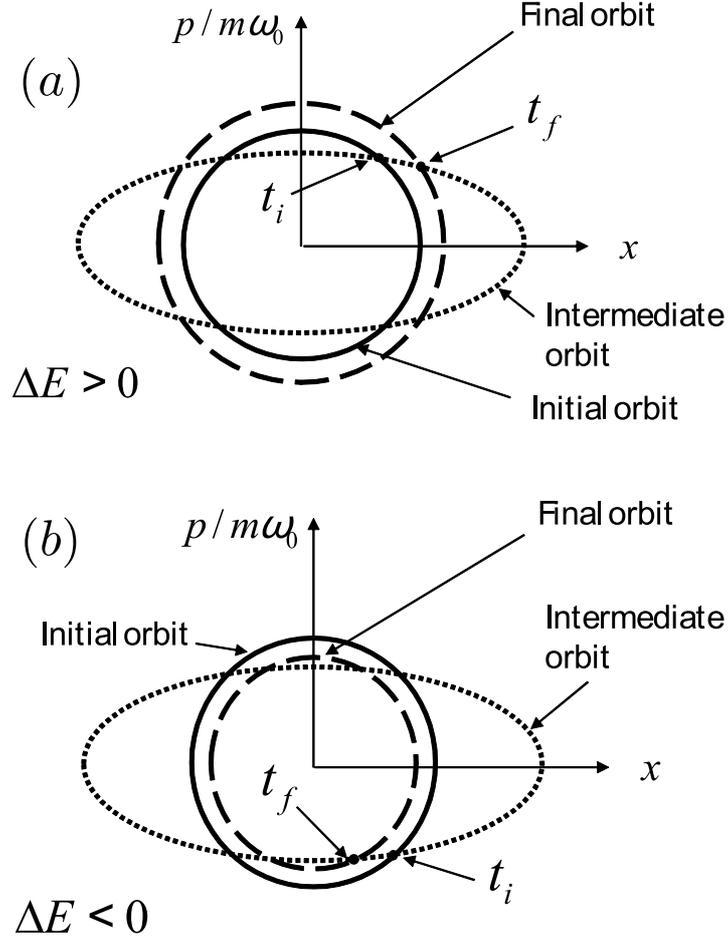}\end{center}

\caption{The system moves clockwise on a circular orbit in phase space until
the spring constant changes at $t=t_{i}$ and it moves onto an elliptical
orbit. The cycle is completed by a change in spring constant at $t=t_{f}$
to return the system to a circular orbit. Depending on the initial
location of the system on its orbit, and the duration of the process,
the radius of the final orbit could be ($a$) greater than or ($b$)
less than the initial radius, illustrating how the energy change can
be both positive or negative. \label{Fig1}}
\end{figure}

It is convenient to convert the time-dependent spring constant $\kappa(t)$
into a time-dependent natural frequency $\omega(t)$, so that the
Hamiltonian at time $t$ is
\begin{equation}
H(t)=p^{2}/2m+m\omega^{2}(t)x^{2}/2,\label{5}
\end{equation}

and the equation of motion is $\ddot{x}=-\omega^{2}x$, where $p$
and $x$ are the momentum and position of the oscillator, and $m$
is its mass. The response of the system may be illustrated using pictures
of phase space orbits. An isolated system with constant frequency
$\omega_{0}$ performs clockwise circular orbits in a phase space
where momentum is normalised by dividing by $m\omega_{0}$. The square
of the radius of the orbit is proportional to the initial energy of
the oscillator. If the spring constant is changed at time $t_{i}$,
altering $\omega$ and the energy, then the system moves onto an elliptical
orbit. A process consisting of a sequence of shifts in spring constant
without contact with a heat bath can therefore be visualised as transitions
to, and movement along, a set of elliptical orbits in phase space.
But if the process is cyclic, characterised by a return to the original
spring constant, the final orbit will be circular. Whether the energy
of the system has increased or decreased as a result of the process
then depends on whether the radius of the final orbit is greater than
or less than the initial radius, respectively. This is illustrated
in Figure \ref{Fig1} for a process consisting of a shift down and
up in spring constant. The implication is that both upward and downward
changes in energy can be generated. With the exception of some special
cases, where motion on the intermediate orbit consists of one complete
circuit for example, in which case $\Delta E=0$ always, a cyclic
process clearly produces a pdf that describes both positive and negative
$\Delta E$.

Let us consider a general variation in $\omega(t)$ in the interval
$0\le t\le\tau$, with $\omega=\omega_{0}=\omega(0)$ for $t<0$ and
$\omega=\omega_{N}=\omega(\tau)$ for $t>\tau$. In the initial and
final situations the motion will be sinusoidal, but with differing
phases and amplitudes in general. We write $x(t)=A\exp(\rmi\omega_{0}t)+A^{*}\exp(-\rmi\omega_{0}t)$,
where $A$ is a complex phasor representing the phase and amplitude.
The process in the interval $0\le t\le\tau$ will then map an initial
phasor $A$ onto a final phasor $A^{\prime}=SA$, with a matrix representation
\begin{equation}
\left(\begin{array}{c}
A_{\rm r}^{\prime}\\
A_{\rm i}^{\prime}
\end{array}\right)=\left(\begin{array}{cc}
a & b\\
c & d
\end{array}\right)\left(\begin{array}{c}
A_{\rm r}\\
A_{\rm i}
\end{array}\right),\label{8}
\end{equation}
where $A_{\rm r}$ and $A_{\rm i}$ are the real and imaginary parts of the
phasor $A$. We shall call $S$ the \emph{process matrix}.

The energy of an oscillation with phasor $A$ and angular frequency
$\omega$ can be written as $2m\omega^{2}A^{T}A$, so the change in
energy brought about by the process is
\begin{equation}
\Delta E=2m\omega_{0}^{2}A^{T}\left(D^{2}S^{T}S-I\right)A,\label{10}
\end{equation}
where $D=\omega_{N}/\omega_{0}$ and $I$ is the unit $2\times2$
matrix. In terms of magnitude $\mid A\mid$ and phase $\theta=\tan^{-1}(\dot{x}(0)/\omega_{0}x(0))$:
\begin{equation}
A=\mid A\mid\left(\begin{array}{c}
\cos\theta\\
\sin\theta
\end{array}\right)\label{11}
\end{equation}
we find that
\begin{equation}
\Delta E=E\left(C_{1}+C_{2}\cos2\theta+C_{3}\sin2\theta\right),\label{12}
\end{equation}
where $E=2m\omega^2_{0}\vert A\vert^{2}$ is the initial energy of the
oscillator, and the constants $C_{1}$, $C_{2}$ and $C_{3}$ can
be expressed in terms of the elements of $S$:
\begin{eqnarray}
& & C_{1}=\frac{D^{2}}{2}\left(a^{2}+b^{2}+c^{2}+d^{2}\right)-1\\
& & C_{2}=\frac{D^{2}}{2}\left(a^{2}-b^{2}+c^{2}-d^{2}\right)\\
& & C_{3}=D^{2}(ab+cd).
\label{13c}
\end{eqnarray}

Let us consider a sequence of step changes in frequency from $\omega_{n-1}$
to $\omega_{n}$ at times $t_{n}=n\delta t$, such that in the interval
$t_{n}\le t\le t_{n+1}$ the displacement is represented by
\begin{equation}
x(t)=A_{n}\exp\left(\rmi\omega_{n}\left(t-t_{n}\right)\right)+A_{n}^{*}\exp\left(-\rmi\omega_{n}\left(t-t_{n}\right)\right).\label{13d}
\end{equation}
Imposing continuity requirements for $x$ and $\dot{x}$ at $t=t_{n}$,
we express $A_{n}$ in terms of $A_{n-1}$:
\begin{eqnarray}
A_{n}
& = &
 \frac{1}{2}\left(A_{n-1}\exp\left(\rmi\omega_{n-1}\delta t\right)\left(1+\frac{\omega_{n-1}}{\omega_{n}}\right)\right.
 \nonumber
\\
 &  &
\left.+A_{n-1}^{*}\exp\left(-\rmi\omega_{n-1}\delta t\right)\left(1-\frac{\omega_{n-1}}{\omega_{n}}\right)\right).\label{14}
\end{eqnarray}
The \emph{step} matrix $S_{n}$, representing the transformation from
$A_{n-1}$ to $A_{n}$ through the relation $A_{n}=S_{n}A_{n-1}$,
has the following structure:
\begin{equation}
\left(\!\begin{array}{cc}
c_{n-1} & -s_{n-1}\\
\frac{\omega_{n-1}}{\omega_{n}}s_{n-1} & \frac{\omega_{n-1}}{\omega_{n}}c_{n-1}
\end{array}\!\right)=\left(\!\begin{array}{cc}
1 & 0\\
0 & \frac{\omega_{n-1}}{\omega_{n}}
\end{array}\!\right)\left(\!\begin{array}{cc}
c_{n-1} & -s_{n-1}\\
s_{n-1} & c_{n-1}
\end{array}\!\right),\label{15}
\end{equation}
where $c_{n-1}=\cos(\omega_{n-1}\delta t)$ and $s_{n-1}=\sin(\omega_{n-1}\delta t)$,
corresponding to a rotation through phase angle $\omega_{n-1}\delta t$,
followed by a rescaling of the imaginary part of the phasor. The process
matrix $S$ may then be constructed from the step matrices as $S=S_{N}S_{N-1}...S_{1}$.
Note that $\det S_{n}=\omega_{n-1}/\omega_{n}$ so that $\det S=(\omega_{N-1}/\omega_{N})\cdots(\omega_{0}/\omega_{1})=\omega_{0}/\omega_{N}=D^{-1}=ad-bc$
and this is unity if the process is cyclic with $\omega_{N}=\omega_{0}$.
By introducing small step changes in $\omega$, we can consider processes
consisting of a continuous variation in spring constant.

\subsection{Construction of $P(\triangle E)$}

Consideration of all possible initial conditions establishes the pdf
of energy change $P(\Delta E)$ due to the process. We assume a canonical
distribution over points $\Gamma$ in phase space now labelled by
initial energy $E$ and phase angle $\theta$:
\begin{equation}
P(\Gamma)\rmd\Gamma=P(\theta\vert E)P(E)\rmd E\, \rmd\theta=\frac{1}{2\pi}\beta \rme^{-\beta E}\rmd E\, \rmd\theta,\label{17}
\end{equation}
and hence the average of $\Delta E$ is
\begin{eqnarray}
& &\langle\Delta E\rangle  \!\!=\!\!\int_{0}^{2\pi}\!\!\int_{0}^{\infty}\!\!\frac{\beta}{2\pi}\rme^{-\beta E}E(C_{1}+C_{2}\cos2\theta+C_{3}\sin2\theta)\rmd E\rmd\theta\nonumber\\
& & =C_{1}/\beta,
\label{eq:18a}
\end{eqnarray}
which, through  (\ref{13c}), establishes a connection between
$\langle\Delta E\rangle$ and the elements of the process matrix $S$.

\begin{figure}
\begin{center}\includegraphics[width=0.7\columnwidth]{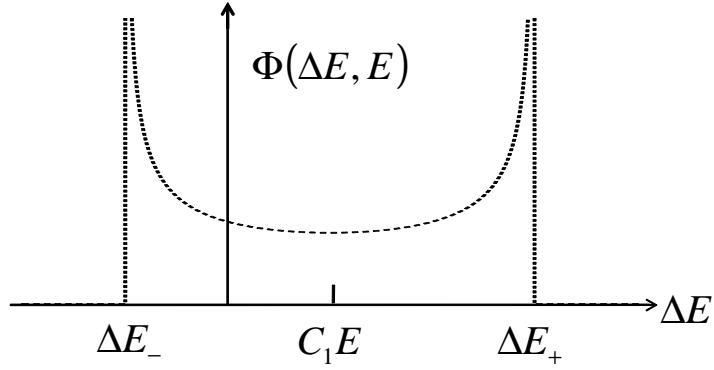}\end{center}

\caption{The pdf of energy change $\triangle E$ for an initial energy $E$,
as specified in (\ref{20}).\label{fig:0} }

\end{figure}

Consider next the pdf $\Phi$ of energy change $\Delta E$ for an
oscillator with a given initial energy $E$. It is straightforward
to write
\begin{eqnarray}
\Phi(\Delta E,E) & = & \sum_{j=1}^{k}P(\theta_{j}\vert E)\left|\frac{\partial\theta}{\partial\Delta E}\right|_{\theta_{j}},\label{18c}
\end{eqnarray}
where ${\theta_{1}...\theta_{k}}$ are the values of $\theta$ that
satisfy (\ref{12}). We employ $\Delta E$ from (\ref{12})
in the following form:
\begin{equation}
\triangle E=E\left(C_{1}+\left(C_{2}^{2}+C_{3}^{2}\right)^{1/2}\cos2(\theta+\theta_{0})\right),\label{18d}
\end{equation}
where $\cos2\theta_{0}=C_{2}/\left(C_{2}^{2}+C_{3}^{2}\right)^{1/2}$.
Noting that, in non-exceptional cases, $k=4$ we write:
\begin{equation}
\Phi(\Delta E,E)=\frac{4}{2\pi}\left|\frac{\partial\theta}{\partial\Delta E}\right|=\frac{1}{\pi\left|E\left(C_{2}^{2}+C_{3}^{2}\right)^{1/2}\sin2\phi\right|},\label{19}
\end{equation}
 where $\phi=\theta+\theta_{0}$, and obtain
\begin{eqnarray}
\Phi(\triangle E,E) & = & \pi^{-1}\left(E^{2}(C_{2}^{2}+C_{3}^{2})-(\triangle E-EC_{1})^{2}\right)^{-1/2}\nonumber \\
 & = & \frac{1}{\pi\left[\left(\triangle E-\Delta E_{-}\right)\left(\triangle E_{+}-\Delta E\right)\right]^{1/2}}.\label{20}
\end{eqnarray}
This is valid for $\triangle E_{-}\le\Delta E\le\triangle E_{+}$,
where $\triangle E_{\pm}=E\left(C_{1}\pm\left(C_{2}^{2}+C_{3}^{2}\right)^{1/2}\right)$.
$\Phi$ is zero outside this range, is symmetric about $\triangle E=EC_{1}$,
and diverges at $\triangle E=\triangle E_{\pm}$, as illustrated in
Figure \ref{fig:0}.

\begin{figure}
\begin{center}\includegraphics[width=0.7\columnwidth]{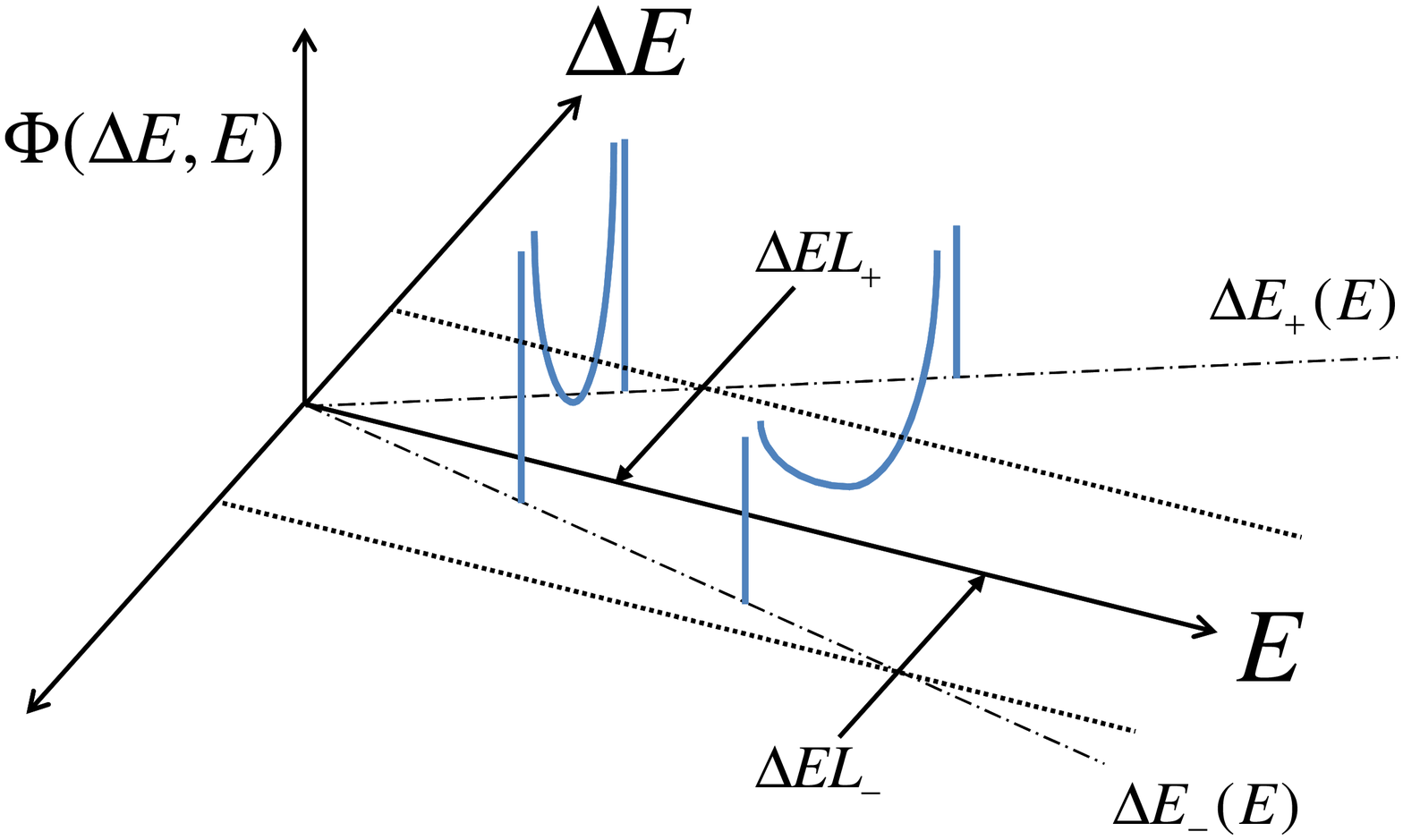}\end{center}

\caption{Indication of the domain of integration over the pdf $\Phi(\Delta E,E)$
for the case $C_{1}>0$. The function itself is sketched by curves
similar to Figure \ref{fig:0} for two values of $E$. \label{fig:domain-of-integration}}

\end{figure}

Now we include the distribution of initial energies $E$. We write
\begin{equation}
P(\triangle E)\propto\int_{0}^{\infty}\Phi(\triangle E,E)\exp(-\beta E)\, \rmd E,\label{21}
\end{equation}
which for $\Delta E\ge0$ may be cast more explicitly as
\begin{equation}
P(\triangle E)\propto\int_{\Delta EL_{+}}^{\infty}\Phi(\triangle E,E)\exp(-\beta E)\, \rmd E,\label{22}
\end{equation}
while for $\Delta E<0$ we use
\begin{equation}
P(\triangle E)\propto\int_{\Delta EL_{-}}^{\infty}\Phi(\triangle E,E)\exp(-\beta E)\, \rmd E,\label{22a}
\end{equation}
where $L_{\pm}=\left(C_{1}\pm\left(C_{2}^{2}+C_{3}^{2}\right)^{1/2}\right)^{-1}$.
We have assumed $C_{2}^{2}+C_{3}^{2}>C_{1}^{2}$ to ensure that both
positive and negative $\Delta E$ are generated by the process: as
we saw earlier this is almost always the case for a cyclic process.
The integration limits are best understood by consideration of Figure
\ref{fig:domain-of-integration}. The lower integration limit $\Delta EL_{+}$
for the $\Delta E\ge0$ case is the energy for which $\Delta E_{+}(E)$,
the upper boundary of the range for which $\Phi(\triangle E,E)$ is
non-zero, is equal to the given $\Delta E$ (shown as the dotted line).
Similarly, the lower integration limit $\Delta EL_{-}$ for the $\Delta E<0$
case is the energy such that $\Delta E_{-}(E)=-\Delta E$.

It is possible to reduce $P(\Delta E)$ to a simple analytic form
for this situation. For positive $\Delta E$ we write
\begin{eqnarray}
P(\Delta E) & \propto & \int_{0}^{\infty}\Phi(\triangle E,E+\triangle EL_{+})\rme^{-\beta(E+\Delta EL_{+})}\, \rmd E\nonumber \\
 & \propto & \rme^{-\beta\Delta EL_{+}}\int_{0}^{\infty}\frac{\exp\left(-\beta E\right)\rmd E}{\left((\Delta E(L_{-}-L_{+})-E)E\right)^{1/2}}\nonumber \\
 & \propto & \rme^{-\beta\Delta E(L_{+}+L_{-})/2}K_{0}\left(\frac{\beta\Delta E}{2}\left(L_{+}-L_{-}\right)\right),\label{eq:100}
\end{eqnarray}
where $K_{0}$ is a modified Bessel function of the second kind. The
pdf for $\Delta E<0$ may be constructed in a similar way:

\begin{eqnarray}
P(\Delta E) & \propto & \int_{0}^{\infty}\Phi(\triangle E;E+\triangle EL_{-})\rme^{-\beta(E+\Delta EL_{-})}\, \rmd E\nonumber \\
 & \propto & \rme^{-\beta\Delta EL_{-}}\int_{0}^{\infty}\frac{\exp\left(-\beta E\right)\rmd E}{\left(-E\left(E-\Delta E(L_{+}-L_{-})\right)\right)^{1/2}}\nonumber \\
 & \propto & \rme^{-\beta\Delta E(L_{+}+L_{-})/2}K_{0}\left(-\frac{\beta\Delta E}{2}\left(L_{+}-L_{-}\right)\right).\quad\label{eq:101}
\end{eqnarray}
We immediately notice a symmetry of the pdf of energy change:
\begin{equation}
P(\Delta E)=\exp(\beta L\Delta E)P(-\Delta E),\label{eq:101a}
\end{equation}
where $L=-(L_{+}+L_{-})$. This is reminiscent of the Crooks relation,
but the distribution on both sides describes a forward process, whereas
the Crooks relation concerns forward and reverse processes. A consequence
is
\begin{eqnarray}
\langle\exp(-\beta L\Delta E)\rangle & = & \int_{-\infty}^{\infty}P(\Delta E)\exp(-\beta L\Delta E)\rmd\Delta E\nonumber \\
 & = & \int_{-\infty}^{\infty}P(-\Delta E)\rmd\Delta E=1,\label{eq:101b}
\end{eqnarray}
and by Jensen's inequality we deduce that $L\langle\Delta E\rangle\ge0$.

We can express $P(\Delta E)$ in terms of process parameters $C_{1}$
and $D$ since $C_{2}^{2}+C_{3}^{2}-C_{1}^{2}=2C_{1}+1-D^{2}$ such
that
\begin{equation}
L_{+}-L_{-}=\frac{-2\left(C_{2}^{2}+C_{3}^{2}\right)^{1/2}}{C_{1}^{2}-C_{2}^{2}-C_{3}^{2}}=\frac{2\left(C_{1}^{2}+2C_{1}+1-D^{2}\right)^{1/2}}{2C_{1}+1-D^{2}},\label{eq:32-1}
\end{equation}
and
\begin{equation}
L_{+}+L_{-}=\frac{2C_{1}}{C_{1}^{2}-C_{2}^{2}-C_{3}^{2}}=-\frac{2C_{1}}{2C_{1}+1-D^{2}},\label{eq:32-2}
\end{equation}
implying that the pdf has a form that depends only on the mean energy
change $\langle\Delta E\rangle=C_{1}/\beta$ and the ratio $D=\omega_{0}/\omega_{N}$,
as long as $C_{2}^{2}+C_{3}^{2}-C_{1}^{2}=2C_{1}+1-D^{2}>0$.

For simplicity let us now focus our attention on a cyclic process
with $\omega_{N}=\omega_{0}$ and hence $D=1$, $L=-(L_{+}+L_{-})=1$
and $L_{+}-L_{-}=(1+2/C_{1})^{1/2}$. The earlier inequality implies
that $\langle\Delta E\rangle\ge0$ and $C_{1}\ge0$. The pdf takes
the form
\begin{equation}
\!\! P(\Delta E)=\frac{\beta\exp\left(\beta\Delta E/2\right)}{\pi\sqrt{2\beta\langle\Delta E\rangle}}K_{0}\left(\!\frac{\beta\vert\Delta E\vert}{2}\left[1+\frac{2}{\beta\langle\Delta E\rangle}\right]^{1/2}\right),\label{eq:33}
\end{equation}
where a normalisation constant has been inserted. This form is consistent
with $\langle\Delta E\rangle=\int_{-\infty}^{\infty}\Delta E\, P(\Delta E)\, \rmd\Delta E$.
It is also consistent with the classical limit of the work distribution
of a quantum harmonic oscillator for a cyclic process obtained by
Deffner and Lutz \cite{Deffner08}.

\begin{figure}
\medskip
\begin{center}\includegraphics[width=0.7\columnwidth]{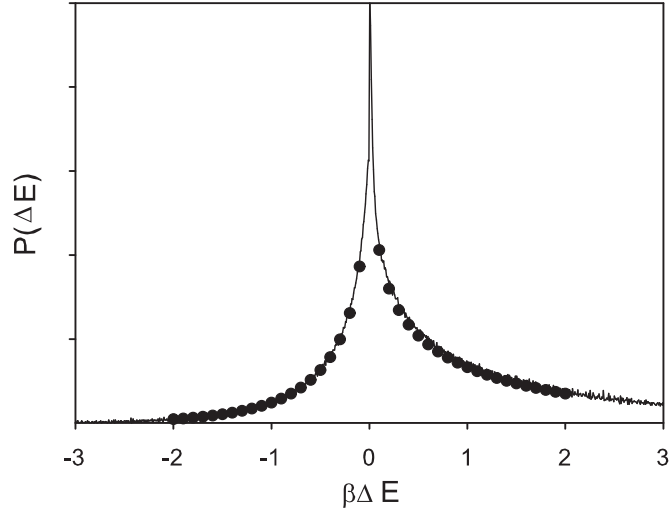}\end{center}

\caption{Distributions of energy change $\Delta E$ produced by a step up and
step down in $\omega$, generated from Monte Carlo (line) and
(\ref{eq:33}) and (\ref{35}) (points) with $\omega_{1}=2\omega_{0}$,
$\omega_{1}\tau=\pi/2$ and $\beta=1$. \label{Fig2}}
\end{figure}

\subsection{Example cases\label{sub:Example-cases}}

The analysis may be illustrated for a cyclic process consisting of
a step change in frequency from $\omega_{0}$ to $\omega_{1}$ at
$t=0$, followed by another jump from $\omega_{1}$ to $\omega_{0}$
at $t=\tau$. The process matrix $S$ takes the form
\begin{equation}
S=\left(\begin{array}{cc}
1 & 0\\
0 & \frac{\omega_{1}}{\omega_{0}}
\end{array}\right)\left(\begin{array}{cc}
c_{1} & -s_{1}\\
s_{1} & c_{1}
\end{array}\right)\left(\begin{array}{cc}
1 & 0\\
0 & \frac{\omega_{0}}{\omega_{1}}
\end{array}\right),\label{34}
\end{equation}
where $c_{1}=\cos\omega_{1}\tau$ and $s_{1}=\sin\omega_{1}\tau$
which implies that
\begin{equation}
C_{1}=\sin^{2}\omega_{1}\tau\left(\frac{1}{2}\left[\left(\frac{\omega_{0}}{\omega_{1}}\right)^{2}+\left(\frac{\omega_{1}}{\omega_{0}}\right)^{2}\right]-1\right).\label{35}
\end{equation}
The results of a Monte Carlo simulation of such a process with $\omega_{1}=2\omega_{0}$
and $\omega_{1}\tau=\pi/2$, with initial states selected from a canonical
distribution with $\beta=1$, are shown as a histogram of energy changes
$\Delta E$ (continuous line) in Figure \ref{Fig2}. The points are
obtained from (\ref{eq:33}) using the appropriate value $C_{1}=9/8$.
The correspondence between the analytical results and the simulation
is apparent.

\begin{figure}
\begin{center}\noindent \includegraphics[width=0.7\columnwidth]{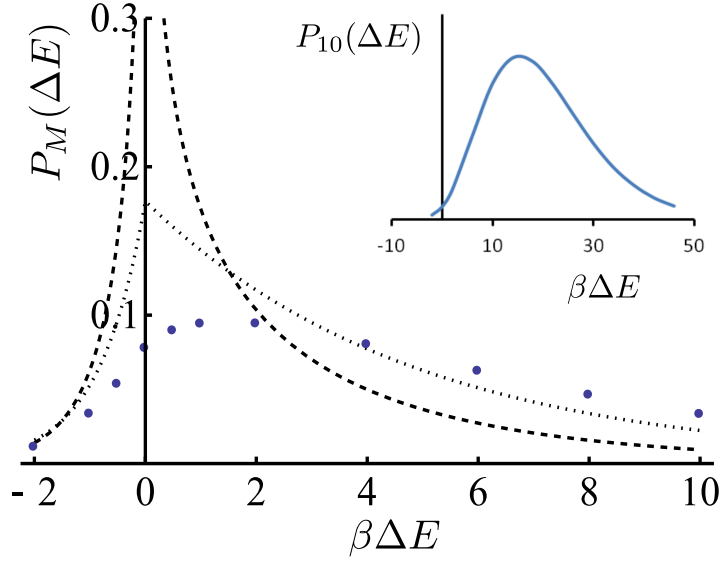}\end{center}

\caption{Probability distribution functions of energy change $\Delta E$ for
ten (solid curve in inset), three (points), two (dotted) and a single
oscillator (dashed), when driven by a work cycle with mean energy
change $\langle\Delta E\rangle=2/\beta$ per oscillator.\label{fig:00}}

\end{figure}

We next calculate $P_{M}(\Delta E)$, the pdf of energy change for
$M$ independent oscillators undergoing a given cyclic process, using
the iterated convolution operation $P_{M}(\Delta E)=\int_{-\infty}^{\infty}P_{1}(\Delta E-x)P_{M-1}(x)\rmd x$,
where $P_{1}(x)$ is synonymous with $P(x)$. Since $P_{2}(-\Delta E)=\int_{-\infty}^{\infty}P_{1}(-\Delta E-x)P_{1}(x)\rmd x=\int_{-\infty}^{\infty}P(-\Delta E+x)P(-x)\rmd x=\int_{-\infty}^{\infty}P(\Delta E-x)\exp(\beta\left(-\Delta E+x\right))P(x)\exp(-\beta x)\rmd x=\exp(-\beta\Delta E)P_{2}(\Delta E)$,
the symmetry of the pdf for a single oscillator is retained for a
system of two oscillators. By iteration, it can be shown that
\begin{equation}
P_{M}(\Delta E)=P_{M}(-\Delta E)\exp(\beta\Delta E).\label{eq:36}
\end{equation}
For a cyclic process with $C_{1}=2$ we show numerically generated
pdfs in Figure \ref{fig:00} for one, two, three and ten oscillators.
All of them satisfy (\ref{eq:36}) in spite of their varying shape
as $M$ increases.

These symmetries of the work distribution, however, are not immediately
equivalent to the Crooks relation, which refers to forward and backward
processes. But it has been remarked that the harmonic oscillator is
rather special \cite{Harris07,Saha09} and that fluctuation relations
of a specific kind emerge. Since the Crooks relation states that $P_{\rm F}(\Delta E)=P_{\rm B}(-\Delta E)\exp(\beta\Delta E)$
for an isolated cyclic process it must be the case that $P_{\rm F}(\Delta E)=P_{\rm B}(\Delta E)$:
the distribution of energy change for the oscillator is the same whether
we process the system according to a forward cyclic sequence $\omega(t)$
or the reverse sequence $\bar{\omega}(t)=\omega(\tau-t)$. The special
nature of the oscillator allows us to understand this in the following
way.

The process matrices for the forward and backward cycles, $S_{\rm F}$
and $S_{\rm B}$ respectively, are related by $S_{\rm F}\hat{S}S_{\rm B}\hat{S}=I$
where
\begin{equation}
\hat{S}=\left(\begin{array}{cc}
1 & 0\\
0 & -1
\end{array}\right)\label{eq:370}
\end{equation}
is a velocity inversion matrix since it transforms phasor $A$ into
$A^{*}$, and these represent oscillator configurations with the same
position but opposite velocities. The relation is simply the statement
that a deterministic forward process, followed by velocity inversion,
the backward process and another velocity inversion, should restore
the initial state of the system in phase space. Hence, if
\begin{equation}
S_{\rm F}=\left(\begin{array}{cc}
a & b\\
c & d
\end{array}\right)\quad\:\mathrm{then\quad}\: S_{\rm B}=\left(\begin{array}{cc}
d & b\\
c & a
\end{array}\right),\label{eq:42-1}
\end{equation}
and clearly, both matrices have the same value of $C_{1}$ according
to (\ref{13c}), in spite of corresponding to quite different
transformations of the initial phasor $A$. We showed in (\ref{eq:33})
that the distribution of energy change is the same for two cyclic
processes with equal values of $C_{1}$ and $\langle\Delta E\rangle$,
and hence $P_{\rm F}(\Delta E)=P_{\rm B}(\Delta E)$.

We noted with reference to Figure \ref{fig:0} that if $C_{2}^{2}+C_{3}^{2}<C_{1}^{2}$
then a distribution with only positive or negative values of $\Delta E$
emerges. For $C_{1}>0$, we find $\Delta E_{-}>0$ and the distribution
would then take the form
\begin{equation}
P(\triangle E)\propto\int_{\Delta EL_{+}}^{\Delta EL_{-}}\Phi(\triangle E,E)\exp(-\beta E)\, \rmd E,\label{eq:300}
\end{equation}
for positive $\Delta E$ and $P(\Delta E)=0$ otherwise, which again
can best be understood with reference to Figure \ref{fig:domain-of-integration}.
However, this does not appear to reduce to a simple form in general
and a symmetry about the point $\Delta E=0$ is obviously absent.
However, a special case for $C_{2}^{2}+C_{3}^{2}=C_{1}^{2}$ does
simplify since it corresponds to $L_{-}\to \infty$ and $L_{+}=1/(2C_{1})$.
We find that
\begin{eqnarray}
& & P(\Delta E)\propto\int_{\Delta EL_{+}}^{\infty}\frac{\rme^{-\beta E}\rmd E}{\left(\triangle E\left(E-\Delta EL_{+}\right)\right)^{1/2}}\nonumber\\
& & \propto\frac{\rme^{-\beta\Delta EL_{+}}}{\Delta E^{1/2}}=\left(\frac{\beta}{2C_{1}\pi\Delta E}\right)^{1/2}\rme^{-\beta\Delta E/(2C_{1})}.
\label{eq:301a-1}
\end{eqnarray}

An example process where this applies is a step up from $\omega_{0}$
to $\omega_{1}>\omega_{0}$: $S$ is simply the right hand component
matrix in (\ref{34}) such that
\begin{equation}
2C_{1}=\left(\frac{\omega_{1}}{\omega_{0}}\right)^{2}\!\left(\!1+\left(\frac{\omega_{0}}{\omega_{1}}\right)^{2}\!\right)-2=\left(\frac{\omega_{1}}{\omega_{0}}\right)^{2}\!-1,\label{eq:301a}
\end{equation}
 and the condition $C_{2}^{2}+C_{3}^{2}-C_{1}^{2}=2C_{1}+1-D^{2}=0$
holds. Thus
\begin{equation}
\! P_{\rm F}(\Delta E)=\left[\frac{\beta\omega_{0}^{2}}{\pi\left(\omega_{1}^{2}-\omega_{0}^{2}\right)\Delta E}\right]^{1/2}\!\!\!\!\!\exp\left(\!-\frac{\beta\omega_{0}^{2}\Delta E}{\omega_{1}^{2}-\omega_{0}^{2}}\right),\label{eq:302}
\end{equation}
for $\Delta E>0$ and zero otherwise. For the reverse process consisting
of a step down from $\omega_{1}$ to $\omega_{0}$ the pdf is
\begin{equation}
\! P_{\rm B}(\Delta E)=\left[-\frac{\beta\omega_{1}^{2}}{\pi\left(\omega_{1}^{2}-\omega_{0}^{2}\right)\Delta E}\right]^{1/2}\!\!\!\!\!\exp\left(\frac{\beta\omega_{1}^{2}\Delta E}{\omega_{1}^{2}-\omega_{0}^{2}}\right),\label{eq:303}
\end{equation}
for $\Delta E<0$ and zero otherwise. The subscripts identify these
as pdfs for a forward process and its backward counterpart, and they
satisfy
\begin{eqnarray}
\frac{P_{\rm F}(\Delta E)}{P_{\rm B}(-\Delta E)} & = & \frac{\omega_{0}}{\omega_{1}}\exp\left(-\frac{\beta\omega_{0}^{2}\Delta E}{\omega_{1}^{2}-\omega_{0}^{2}}\right)\exp\left(\frac{\beta\omega_{1}^{2}\Delta E}{\omega_{1}^{2}-\omega_{0}^{2}}\right)\nonumber \\
 & = & \exp(\beta(\Delta E-\Delta F)),\label{eq:304}
\end{eqnarray}
as required by the Crooks relation, where $\Delta F=\beta^{-1}\ln(\omega_{1}/\omega_{0})$
is the free energy change in the forward process. They resemble the
work distributions for a harmonic oscillator under isothermal conditions
for this process \cite{SpinneyFordChapter12}.

\begin{figure}
\begin{center}\includegraphics[width=0.7\columnwidth]{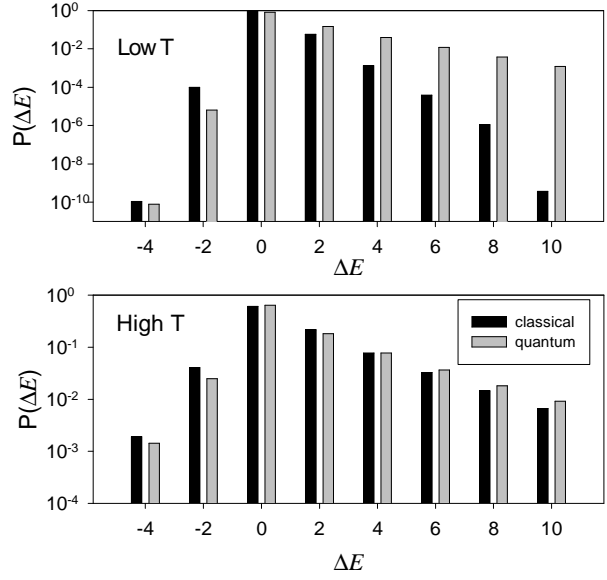}\end{center}

\caption{Probabilities of energy change $\Delta E$ (in units of $\hbar\omega_{0}$)
for a step up-step down cyclic work process. Grey bars represent $P_{\rm q}(\Delta E)$
in (\ref{eq:38a}), derived from a quantum treatment, and black
bars correspond to the classical pdf (\ref{eq:33}), integrated
over a range $\hbar\omega_{0}$ each side of the allowed quantised
values of $\Delta E$. The lower plot at $\beta\hbar\omega=1$ illustrates
the similarity of treatments at high temperature in contrast to the
upper plot at $\beta\hbar\omega=5$ (or low temperature) where the
range of fluctuations is broader in the quantum treatment compared
with the classical. The quantum histograms satisfy $P_{\rm q}(\Delta E)=\exp(\beta\Delta E)P_{\rm q}(-\Delta E)$:
the classical counterparts do not though they are based on underlying
continuous pdfs that do.\label{fig:3}}
\end{figure}

\section{Quantum treatment\label{sec:Quantum-treatment}}

We expect the pdfs of energy change for a quantum treatment of forward
and backward processes for an isolated oscillator to satisfy the
Crooks relation, as has been demonstrated by Deffner \emph{et al}
\cite{Deffner08,Deffner10}. However, our interest in this section
is in the symmetry in the distribution of energy change for a forward
cyclic process. We employ the treatment of a 1-d quantum oscillator
driven by an arbitrary $\omega(t)$ provided by Ji \emph{et al} \cite{Ji95}.
According to this approach, the familiar eigenfunctions $\psi_{n}(x)$
of a 1-d harmonic oscillator with frequency $\omega(0)=\omega_{0}$
evolve into
\begin{eqnarray}
& & \!\hat{\psi}_{n}(x,t)=\frac{1}{2^{n}n!}\left(\frac{\omega_{0}}{\hbar\pi g_{-}(t)}\right)^{\nicefrac{1}{4}}\!\exp\left[-\frac{\left(\omega_{0}+\rmi g_{0}(t)\right)}{2\hbar g_{-}(t)}x^{2}\right.\nonumber\\
& & \left.-\rmi\left(n+\frac{1}{2}\right)\int_{0}^{t}\frac{\omega_{0}}{mg_{-}(t^{\prime})}dt^{\prime}\right]H_{n}\left(\sqrt{\frac{\omega_{0}}{\hbar g_{-}(t)}}x\right),
\label{eq:37a}
\end{eqnarray}
where the $H_{n}$ are Hermite polynomials and the functions $g_{-}(t)$,
$g_{0}(t)$ and $g_{+}(t)$ satisfy
\begin{eqnarray}
\dot{g}_{-}(t) & = & -2g_{0}(t)/m\nonumber \\
\dot{g}_{0}(t) & = & m\omega^{2}(t)g_{-}(t)-g_{+}(t)/m\nonumber \\
\dot{g}_{+}(t) & = & 2m\omega^{2}(t)g_{0}(t),
\end{eqnarray}
with initial conditions $g_{-}(0)=1/m$, $g_{0}(0)=0$ and $g_{+}(0)=m\omega_{0}^{2}$.
The pdf of energy change $P_{\rm q}(\Delta E)$ for an arbitrary cyclic
process is then straightforward to calculate. We write
\begin{equation}
P_{\rm q}(j\hbar\omega_{0})=\sum_{n=0}^{\infty}\sum_{k=0}^{\infty}\left|T_{kn}(t)\right|^{2}p_{n}\delta_{k-n,j},\label{eq:38a}
\end{equation}
where $p_{n}\propto\exp(-\left(n+\nicefrac{1}{2}\right)\hbar\omega_{0}\beta)$
is the initial canonical probability for state $n$, and $T_{kn}(t)$
is the transition amplitude, which up to an unimportant phase is given
by
\begin{eqnarray}
& & \!\! T_{kn}(t)=\int_{-\infty}^{\infty}dx\frac{1}{2^{k}k!}\left(\frac{\omega_{0}}{\hbar\pi g_{-}(t)}\right)^{\nicefrac{1}{4}}\nonumber\\
& & \times\exp\left[\frac{\left(\rmi g_{0}(t)-\omega_{0}\right)}{2\hbar g_{-}(t)}x^{2}\right]H_{k}\left(\sqrt{\frac{\omega_{0}}{\hbar g_{-}(t)}}x\right)\nonumber\\
& & \times\frac{1}{2^{n}n!}\left(\frac{m\omega_{0}}{\hbar\pi}\right)^{\nicefrac{1}{4}}\exp\left[-\frac{m\omega_{0}}{2\hbar}x^{2}\right]H_{n}\left(\sqrt{\frac{m\omega_{0}}{\hbar}}x\right).
\label{eq:41f}
\end{eqnarray}
Parity considerations dictate that $T_{kn}$ is zero unless $k$ and
$n$ differ by an even number, and hence $P_{\rm q}(j\hbar\omega_{0})$
is zero unless $j$ is even.

We study the step up and down process that was considered classically
in section \ref{sub:Example-cases}, namely $\omega=\omega_{0}$ for
$t<0$ and $t>\tau$, and $\omega=\omega_{1}=2\omega_{0}$ in the
interval $0\le t\le\tau$ with $\omega_{1}\tau=\pi/2$. The pdfs of
energy change under the quantum dynamics for cases where $\beta\hbar\omega_{0}$
is equal to 1 and 5 are shown in Figure \ref{fig:3}. Both pdfs satisfy
the relationship $P_{\rm q}(\Delta E)=\exp(\beta\Delta E)P_{\rm q}(-\Delta E)$.
The quantum pdfs may be contrasted with the classical counterparts
by integrating $P(\Delta E)$ in (\ref{eq:33}) over the range
$(j-1)\hbar\omega_{0}\le\Delta E\le(j+1)\hbar\omega_{0}$ and comparing
the result with $P_{\rm q}(j\hbar\omega_{0})$. We see that the classical
and quantum treatments coincide rather well for $\beta\hbar\omega_{0}=1$,
but that differences emerge for $\beta\hbar\omega_{0}=5$. The latter
is evidently a low temperature regime and the energy change brought
about by the process is distributed more broadly due to the relatively
more substantial quantum fluctuations.

\begin{figure}
\begin{center}\includegraphics[width=0.7\columnwidth]{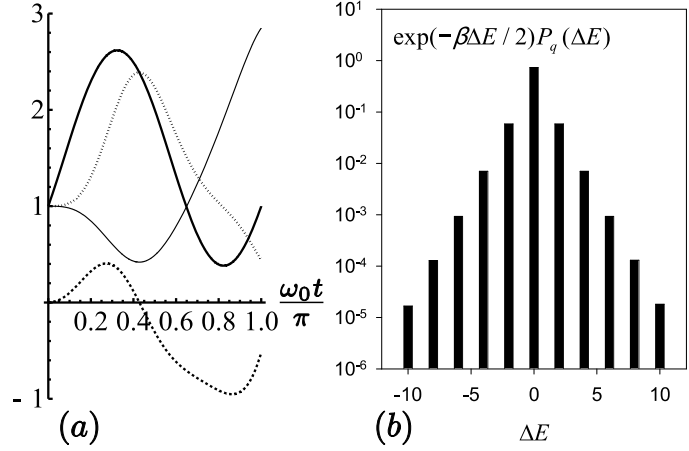}\end{center}

\caption{($a$) The evolution of functions $mg_{-}(t)$ (thin solid curve),
$g_{0}(t)$ (dashed) and $g_{+}(t)/m\omega_{0}^{2}$ (dotted) under
time-dependent frequency driving $\omega(t)/\omega_{0}=\left(1+\sin(2\omega_{0}t)+\sin^{2}(\omega_{0}t)\right)$
(thick solid curve). ($b$) The probability distribution $P_{\rm q}(\Delta E)$
of energy change $\Delta E$ (in units of $\hbar\omega_{0}$) for
$\beta\hbar\omega_{0}=1$ is plotted in the form $f(\Delta E)=\exp(-\beta\Delta E/2)P_{\rm q}(\Delta E)$
to demonstrate the symmetry $f(\Delta E)=f(-\Delta E)$. \label{fig:4}}

\end{figure}

The step up-step down process has a very simple time-dependence of
$\omega(t)$. We investigate driving the system with the more complicated
frequency history $\omega(t)=\omega_{0}\left(1+\sin(2\omega_{0}t)+\sin^{2}(\omega_{0}t)\right)$
over the interval $0\le\omega_{0}t\le\pi$ such that the $g$-functions
evolve with time as in Figure \ref{fig:4}($a$). The resulting distribution
$P_{\rm q}(\Delta E)$ for $\beta\hbar\omega_{0}=1$ is shown in Figure
\ref{fig:4}($b$) in a form that demonstrates the symmetry $P_{\rm q}(\Delta E)=\exp(\beta\Delta E)P_{\rm q}(-\Delta E)$.

\section{Conclusions\label{sec:Conclusions}}

We have calculated the probability distribution function $P(\Delta E)$
of energy change brought about by taking a 1-d harmonic oscillator,
initially in thermal equilibrium, through a process of external work
while isolated from the environment. If the process is cyclic, then
the pdf extends over positive and negative $\Delta E$ and exhibits
a symmetry $P(\Delta E)=\exp(\beta\Delta E)P(-\Delta E)$ with respect
to the reversal of the sign of the energy change. This is reminiscent
of the Crooks relation, which is also satisfied by the system, but
is distinct, since it involves a forward process only and not its
reverse. But the symmetry is also a special case of the result $P(\Delta E)=\exp(\beta L\Delta E)P(-\Delta E)$
that is valid for a class of non-cyclic processes for this system.
These symmetries are a consequence of the simple dynamics of an oscillator.
We have demonstrated that the symmetry is retained if the system consists
of $M$ independent oscillators subjected to the same process, even
though the $P_{M}(\Delta E)$ take a variety of forms. The symmetry
is also retained when an oscillator undergoing a cyclic process is
treated quantum mechanically. At high temperatures the results of
the classical and quantum treatments are similar, while at low temperatures
there is a relative broadening of the pdf in the quantum treatment,
as would be expected from the inclusion of quantum uncertainty in
the dynamics.

The understanding of nonequilibrium thermodynamic processes has advanced
tremendously in the last decade or so as a result of the development
of fluctuation relations and particular identities such as the Crooks
relation and the Jarzynski equality \cite{Harris07,Campisi11}. The
explicit calculation of probability distribution functions satisfying
these relations is a challenging task, but has considerable pedagogical
value. The harmonic oscillator has been a popular system for such
activity, but it is a rather special case in that further symmetries
emerge that are not present in general. Using an approach based on
phasors and a geometrical consideration of phase space trajectories
and their weighting in canonical averages, we have demonstrated some
of this richness. It offers a contrast to approaches offered elsewhere
\cite{Deffner08,Saha09,vanZon08a,Huber08,Deffner10} with the intention
that it might make an additional contribution to this understanding.

\section*{References}

\end{document}